\documentclass[prl,showpacs,preprintnumbers,twocolumn]{revtex4}
\usepackage{graphicx,epsfig}
\usepackage{amssymb,amsmath,amsthm}

\begin{document}
\title{Analytic Formulation, Exact Solutions, and Generalizations of the \\
Elephant and  the Alzheimer Random Walks}
\author{ V. M. Kenkre}
\affiliation{Consortium of the Americas for Interdisciplinary
Science and Department of Physics and Astronomy, University of New
Mexico, Albuquerque, New Mexico 87131, USA}

\begin{abstract}
An analytic formulation of memory-possessing random walks introduced recently [Cressoni et al., Phys. Rev. Lett. 98, 070603 (2007) and Sch\"utz and Trimper, Phys. Rev. E 70, 045101 (2004)] for Alzheimer behavior and related phenomena is provided along with exact solutions  on the basis of Fokker-Planck equations. The solution of a delay-differential equation derived for the purpose is shown to produce log-periodic oscillations and to coincide rather accurately with previously published computer simulation results. Generalizations along several directions are also constructed on the basis of the formalism.
\end{abstract}


\maketitle

Two remarkable publications have recently appeared on the subject of random walks with memory, one involving the `elephant walk' in which the walker chooses steps randomly but is influenced by a perfect memory of steps taken earlier \cite{st}, and the other involving an extension of this walk to incorporate partial memory of steps from the beginning up to a time in the past \cite{cressPRL}. While an analytical description has been given for the (former) elephant walk, it appears to have been impossible to provide one for the partial memory extension.  The significance of the latter is that it has been proposed \cite{cressPRL}  for the medically important analysis of amnestically induced behavior of Alzheimer patients. The authors of Ref. \cite{cressPRL} have presented impressive computer simulations of the Alzheimer walk  exhibiting log-periodic oscillations in the displacement of the walker, and deduced intriguing conclusions regarding the elements of persistence and what they have called, following Sch\"utz and Trimper \cite{st},  traditional versus reformer behavior of the walker. They have also stated that an analytic solution of their partial memory extension (the Alzheimer walk) remains an open problem. The present Letter is aimed at solving that problem.

As a result of both its complexity and its usefulness, the general subject of random walks has occupied exceptional minds for a long time whether in physics \cite{einstein} or in other areas of research \cite{bach}. Contributions in this topic and related areas have continued to be made in physical \cite{montroll,katja,mike,vmkgme,grigo} as well as interdisciplinary sciences \cite{bruce, metzler,mantegna, gandhi} such as sociology, economics and ecology. Progress in the subject has, obviously, multiple beneficial effects.

Let us restrict the analysis to 1-d for simplicity. The original elephant walk \cite{st} envisages a walker that initially starts out in one direction, remembers its entire history at any time, selects with uniform probability any of the steps taken in the past and then imitates that step with probability $p$ or takes the reverse step with probability $1-p$ where $0<p<1.$ The extension introduced by Cressoni et al. \cite{cressPRL} is to restrict the memory of the walker to a fraction $f$ of the total time $t$ elapsed. While this single change results, as shown in Ref. \cite{cressPRL}, in noteworthy consequences such as log-periodic oscillations, it also makes analytic understanding of the evolution difficult.

The present analysis is based on a Fokker-Planck equation, in continuous time $t$ and space $x$, for the probability density $P(x,t)$ of the walker, arrived at by combining the standard continuity equation for $P(x,t)$ and a constitutive equation relating the latter to the current density $j(x,t)$. The constitutive equation is determined by the rules of the random walk. Let us first attempt the form
\begin{equation}
j(x,t)=\frac{\alpha}{ft}\left[xP(x,ft)\right]-D\frac{\partial P(x,t)}{\partial x}
\label{j}
\end{equation}
following the notation of Ref. \cite{st} rather than of Ref. \cite{cressPRL}: the symbol $\alpha$ represents $2p-1$ rather than the Hurst exponent. The resulting Fokker-Planck equation (FPE) would then be
\begin{equation}
\frac{\partial P(x,t)}{\partial t}=-\frac{\alpha}{ft}\frac{\partial}{\partial x}[xP(x,ft)]+D\frac{\partial ^2 P(x,t)}{\partial x^2}
\label{fpe}
\end{equation}
 and reduce in its perfect memory version ($f=1$) to the FPE given as a large $t$ and $x$ limit in Ref. \cite{st}. The second term in each of equations (\ref{j}) and (\ref{fpe}) refers to the random element in the walk whereas the first term expresses the fact that the change in the displacement (the next step taken) at time $t$ is related to an average with uniform weight of all earlier displacements until $ft$, the $ft$ in the denominator being a normalizing factor.

Multiplying (\ref{fpe}) by $x$ and integrating, we get for the first moment $\overline{x}(t)=\int_{-\infty}^{\infty} xP(x,t)dx$
\begin{equation}
\frac{d\overline{x}(t)}{dt}=\frac{\alpha}{ft} \overline{x}(ft).
\label{panto}
\end{equation}
This explicit equation for the average displacement of the walker may be also obtained by introducing the partial memory suggestion of Ref. \cite{cressPRL} into the elephant walk analysis of Ref. \cite{st}, particularly equation (4) of the latter. Let us now solve equation (\ref{panto}) to \emph{demonstrate} that the  average displacement can exhibit logarithmic oscillations acquiring negative as well as positive values, precisely as in the computer simulations of Cressoni et al. \cite{cressPRL}. To this end,  (\ref{panto}) may be transformed through the introduction of a dimensionless time $T$,
\begin{equation}
T=\frac{\ln{(t/t_0)}}{\ln{(1/f)}},
\label{transfo}
\end{equation}
into the delay-differential equation 
\begin{equation}
\frac{dz(T)}{dT}=Bz(T-1)
\label{dde}
\end{equation}
where $z(T)\equiv \overline{x}[t_0(1/f)^T]$ is the displacement expressed as a function of the transformed time $T$, and $B$, which can be negative  for reformers ($p<1/2$), is given by $(\alpha /f) \ln{(1/f)}$. General theorems \cite{theorem} of delay-differential equations may be applied to determine the behavior. Thus, positive $\alpha$ will always lead to monotonic increase of the displacement but negative $\alpha$ may lead to oscillations depending on its value. In the pre-transformation variable $t$, these appear as log-periodic oscillations provided $-B>1/e$. The present theoretical analysis shows that the curve $$(1-2p) \ln{(1/f)}=f/e$$ provides the essence of the $f-p$ phase diagram  mentioned by Cressoni et al. \cite{cressPRL} on the basis of their simulations as separating regions displaying different behavior.

Explicit solutions of (\ref{dde}) may be obtained via Laplace transforms but require specification of initial conditions not only at a single point but over a segment. The moment is naturally taken to be vanishing for all times before $t_0$ and at that time to have the value $x_0$ in keeping with the problem as stated in refs. (\cite{st,cressPRL}). The final solution of (\ref{dde}), when transformed into the $t$-domain, gives
\begin{equation}
\frac{\overline{x}(t)}{x_0}=\mu(t)=\sum_{r=0}^{\infty} \frac{\left[\frac{\alpha}{f}\ln\frac{1}{f}\left(\frac{\ln(t/t_0)}{\ln(1/f)}-r\right)\right]^r}{r!}\Theta\left(t-t_0(1/f)^r\right)
\label{solnback}
\end{equation}
where $\Theta$ is the Heaviside step function--it naturally appears  because of  the delay nature of the equation.

Equation (\ref{solnback}), one of the main results of this Letter,  may be considered  the solution of the open problem mentioned by Cressoni et al \cite{cressPRL}. Figure 1, drawn in close similarity to Fig. 2 of Ref. \cite{cressPRL} compares directly the simulation results of Cressoni et al. taken as data\cite{notte} with our theoretical result (\ref{solnback}).  
\begin{figure}[t] 
   \centering
   \includegraphics[width=3.5in,height=3in]{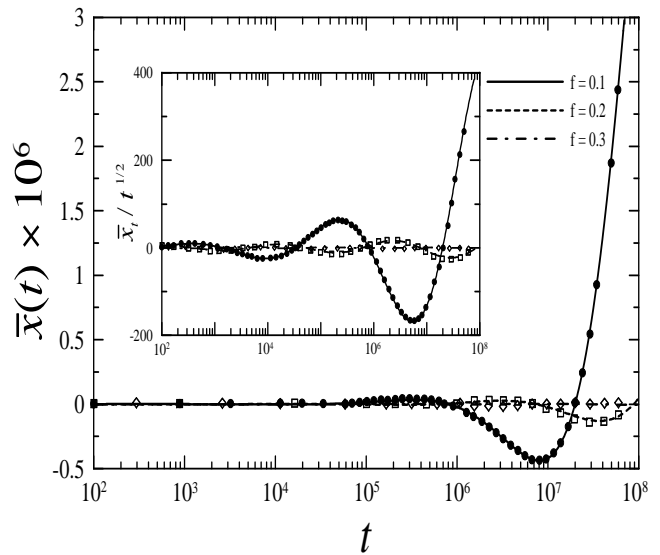} 
   \caption{Comparison of the analytic result (\ref{solnback}) derived in the present paper with the computer simulations of Ref. \cite{cressPRL} showing excellent agreement. Plotted is the average displacement of the walker for the partial memory walk as a function of time for $p=0.1$ and  various values of  $f$, the fraction of the past that the walker remembers. Remarkably, the analytic curve coincides exactly with the simulation data over 6 orders of magnitude while the data executes log-periodic oscillations over the wide range shown.}
   \label{Fig1}
\end{figure}

The amazing coincidence of the analytical expression $\mu(t)$ derived in this paper with the simulation results given in Ref. \cite{cressPRL} should leave little doubt as to the correctness of the claim that the problem, or at least those parts of it addressed by the first moment of the probability distribution, has been indeed solved analytically. Notice that $\overline{x}(t)$ executes logarithmic oscillations of enormous amplitude over many orders of magnitude in time for negative $\alpha$ ($p<1/2$) and yet our analytic curves follow the simulation data faithfully and accurately over the entire range.

The success of the above analysis is marred by the fact that proceeding with higher moments from (\ref{fpe}) lands one in problems. The second moment $\overline{x^2}(t)=\int_{-\infty}^{\infty} \overline{x^2} P(x,t)dx$, when transformed   into $y(T)\equiv \overline{x^2}[t_0(1/f)^T]$, produces
\begin{eqnarray}
y(T)=y(0)\sum_{r=0}^{\infty} \frac{\left[2B(T-r)\right]^r}{r!}\Theta(T-r)+2Dt_0\sum_{r=0}^{\infty}&&\nonumber \\ (2\alpha) ^r f^{-T}\gamma \left(r+1,(T-r)\ln \frac{1}{f}\right)\Theta(T-r),
\label{soln2}
\end{eqnarray}
$\gamma$ being the incomplete gamma function. The right hand side of (\ref{soln2}) can go negative for certain parameter values. What this means is that the FPE (\ref{fpe}), although it gives the correct first moment, predicts probabilities that can go negative. An alternative FPE must be found which does not suffer from this shortcoming but which continues to predict the first moment correctly. One way to do this is to modify the constitutive relation (\ref{j}) by taking in its first term the probability density $P(x,t)$ at the present time rather than at $ft$ and to regard the $x$ that multiplies it as the location in space which, if occupied by the walker at time $ft$ would bring the walker to $x$ at time $t$. Because we have already been able in this paper to solve for the trajectory equation, that is, for the motion of the walker in the absence of the random element ($D=0$), the solution being in our equation (\ref{solnback}), this modification of the constitutive relation is possible to implement. Regarding $x(t)$ as a trajectory variable (no bars over the symbol), we know it explicitly as $x_0\mu(t)$ in terms of our solution. Therefore we also know it at time $ft$ in the past: it is  $x_0\mu(ft)$. The latter quantity may then be expressed in terms of the former quantity via the ratio $\mu(ft)/\mu(t)$. The location which, if occupied by the walker at time $ft$ would bring the walker to $x$ at time $t$, is thus given by the product of the present location $x$ and  $\mu(ft)/\mu(t)$. Substituting this in the constitutive relation, the new form of the relation (to replace (\ref{j}))  becomes
\begin{equation}
j(x,t)=\frac{\alpha}{ft}\frac{\mu(ft)}{\mu(t)}\left[xP(x,t)\right]-D\frac{\partial P(x,t)}{\partial x}
\label{jj}
\end{equation}
the $x$'s in this equation being all regular variables. All the ingredients of `$x$ at time $ft$' have been moved to the time functions. These time functions can be simplified as a result of (\ref{panto}) regarded as a trajectory equation.

Finally, making the simplifications, we obtain, for the correct description of the first as well as the second moments, a new form for the constitutive relation and the FPE. With $h(t)=\frac{d\ln[\mu(t)] }{dt}$,  we have
\begin{equation}
j(x,t)=h(t)\left[xP(x,t)\right]-D\frac{\partial P(x,t)}{\partial x}
\label{j2}
\end{equation}
and
\begin{equation}
\frac{\partial P(x,t)}{\partial t}=-h(t)\frac{\partial}{\partial x}[xP(x,t)]+D\frac{\partial ^2 P(x,t)}{\partial x^2}.
\label{fpe2}
\end{equation}
They lead to the mean square displacement 
\begin{equation}
\overline{x^2}(t)=(\overline{x}(t))^2+2D\mu^2(t)\int_{t_0}^t\frac{ds}{\mu^2(s)}
\label{msd}
\end{equation}
which is guaranteed to be non-negative. Both the terms in the right hand side are already available in the calculations in (\ref{solnback}). A comparison of this expression with the simulation data of Cressoni et al. \cite{cressPRL} seems to show reasonable agreement in both the oscillations (when they are present) and the overall behavior. These and other related matters will be the subject of a future publication \cite{joneskenkre}.

Thus, by substituting (\ref{j2}) and (\ref{fpe2}) for their earlier versions, one is able to get accurate and consistent results for all moments (the first and the second moment being in (\ref{solnback}) and (\ref{msd})). Indeed, using methods explained elsewhere \cite{sevillakenkre} in  context of nuclear magnetic resonance microscopy, an explicit solution for $P(x,t)$ can be obtained. In the Fourier domain, the solution is 
\begin{equation}
\hat{P}(k,t)=\hat{P}(ke^{\int_{t_0}^t h(s)ds},{t_0}) e^{-\left[Dk^2\int_{t_0}^t dt'e^{2\int_{t'}^t h(s)ds}\right]}
\label{fouriersoln}
\end{equation}
The first factor on the right hand side is obtained by replacing $k$ in $\hat{P}(k,{t_0})$, the Fourier transform of the initial probability density, by $ke^{\int_{t_0}^t h(s)ds}$. Inversion of (\ref{fouriersoln}) is trivial\cite{sevillakenkre}.

Next, let us consider imperfect memory generalizations of the elephant walk not in the sense that the walker has a restriction of the time for which it remembers (as in Ref. \cite{cressPRL}) but to describe the situation in which the walker selects one of the  previous steps with some \emph{given} weight depending on when that step was taken. This weight will be described by a function $a(t)$ instead of the constant 1. The integral of this $a(t)$, i.e., $\int_0^t a(s) ds$, will be called $b(t)$: note that b(0)=0. While in Ref. \cite{st} the description is discrete-time, one may write for the continuous time description employed in this Letter, in contrast to their equation (4),
$x(t)=x_0+\int_0^t \sigma(s) ds$,
where $x$ is the position and $\sigma$ is the displacement (step taken at time t).  Generalizing via weights $a(t)$ leads to
\begin{equation}
\frac{dx(t)}{dt}=\sigma(t)=\alpha \left(\frac{\int_0^t a(s)\sigma(s) ds}{\int_0^t  a(s)ds}\right).
\end{equation}
The integral of $a(t)$ appears in the denominator for normalization. Using $a(t)=db(t)/dt$, one obtains
\begin{equation}
\sigma(t)b(t)(1-\alpha)=-\alpha \int_0^t b(s)\frac{d\sigma(s)}{ds} ds,
\end{equation}
and through further differentiation, into
\begin{equation}
\frac{d \ln {\sigma(t)}}{dt}+(1-\alpha)\frac{d \ln {b(t)}}{dt}=0.
\end{equation}
This leads to
\begin{equation}
\frac{dx}{dt}=\frac{[b(t)]^{\alpha-1}}{\int_0^t [b(s)]^{\alpha-1} ds} [x(t)-x(0)],
\label{three}
\end{equation}
to the FPE
\begin{equation}
\frac{\partial P(x,t)}{\partial t}=-h(t) \frac{\partial \left[x P(x,t)\right]}{\partial x}+D\frac{\partial ^2 P(x,t)}{\partial x^2}.
\label{fpgeneralized}
\end{equation}
and to the displacement (first moment) equation
\begin{equation}
\frac{d\overline{x}(t)}{dt}=h(t) \overline{x}(t)
\label{three}
\end{equation}
where the function $h(t)$ is given by
\begin{equation}
h(t)=\frac{d}{dt}\left[\ln{\int_0^t [b(s)]^{\alpha-1} ds}\right].
\label{h}
\end{equation}

This generalization of the elephant walk can be solved explicitly for arbitrary weights $a(t)$ for  which $h(t)$ can be computed. Indeed, this has already been formally done above in (\ref{fouriersoln}). One can write solutions of the FPE and also obtain, and solve, if necessary, arbitrary moment equations. As an example, note that for  $a(t)=A_nt^n$ where $A_n$ is a constant, little seems to change relative to the simple elephant walk case $n=0$ except that the coefficient $\alpha$ is replaced by $(n+1)\alpha-n$. Nevertheless, it should be noticed that the change is not trivial. With $p$ lying between 1/2 and 1, or $\alpha=2p-1$ lying between 0 and 1 (probability of rightward march), $(n+1)\alpha-n$ can become negative. Reformer behavior will then occur even for a positive $\alpha$.

It is also instructive to see how the original results of Sch\"utz and Trimper \cite{st} can be obtained straightforwardly from our formalism. If we take the governing equation to be (\ref{fpe2}) with $h(t)=\alpha/t$, the \emph{complete} solution is in our expression (\ref{fouriersoln}). Inverting the Fourier transform, one obtains for $P(x,t)$, \cite{note2} 
\begin{equation}
P(x,t)=\frac{e^{-\frac{[x-x_0\xi(t)]^2}{4D\eta(t)}}}{\sqrt{4\pi D \eta(t)}},
\label{pdf}
\end{equation}
the first moment being given by
\begin{equation}
\overline{x}(t)=x_0\xi(t),
\label{xbar}
\end{equation}
$\xi$ playing the same role here as $\mu$ in the imperfect memory case. The second moment is
\begin{equation}
\overline{x^2}(t)=[x_0\xi(t)]^2+2D\eta(t).
\label{xsquaredbar}
\end{equation}
The functions $\xi(t)$ and $\eta(t)$ in these formulae may be computed for arbitrary $h(s)$ via
\begin{eqnarray}
\xi(t)&=&e^{\int_{t_0}^t ds\, h(s)},\\
\eta(t)&=&\int_{t_0}^t dt'\, e^{2\int_{t'}^t ds\, h(s)}=\int_{t_0}^t ds\, \frac{\xi^2(t)}{\xi^2(s)}.
\label{fns}
\end{eqnarray}
but reduce, for the elephant walk, to $\xi(t)=(t/t_0)^\alpha$ always, 
and to $\eta(t)=\frac{t-t_0(t/t_0)^{2\alpha}}{1-2\alpha}$ unless $\alpha=1/2$ in which case $\eta(t)=t\ln{(t/t_0)}$. We see that for sufficiently large times the larger power of $t$ dominates and so, one gets the interesting \emph{transition} pointed out by Sch\"utz and Trimper \cite{st}. For $\alpha < 1/2$, we have a linear $t$-behavior of $\eta$ whereas for $\alpha > 1/2$, $\eta$ goes as $t^{2\alpha}$. This is also seen in the mean square displacement:
\begin{equation}
\overline{x^2}(t)=x^2_0\left(\frac{t}{t_0}\right)^{2\alpha}+\frac{2Dt_0}{1-2\alpha}\left[\frac{t}{t_0}-\left(\frac{t}{t_0}\right)^{2\alpha}\right].
\end{equation}
Many other results from Ref.\cite{st} can be rederived in this manner but will not be shown for want of space. 


In summary, this Letter has provided an analytic formulation and exact solutions for memory-possessing random walks exhibiting novel effects such as persistent behavior and log-periodic solutions formerly found only via computer simulations. Whether or not Alzheimer walks as proposed by Cressoni et al. \cite{cressPRL} will have medical significance, they are certainly interesting random walks in their own right. It is hoped that this work has made a contribution towards their understanding. We note in passing that here we have refrained from introducing memory functions into the diffusion term (the last term of the Fokker-Planck equation) which is typically the place memories have been often introduced in the past to study coherence \cite{vmkgme} turning diffusion equations into telegrapher's equations and including thus wave effects. That is a quite different family of effects from those studied in refs. \cite{st,cressPRL} or here. 
 
It is a pleasure to thank Dara Jones for discussions and assistance including  in the preparation of the plots; also Ben Baragiola, Colston Chandler and Yiu-Man Wong  for a number of insightful conversations. This work has been supported in part by the NSF's INT-0336343, DARPA-N00014-03-1-0900 and NIH/NSF's EF-0326757.


\begin{references}
\bibitem{st} G.M. Sch\"utz and S. Trimper, Phys. Rev. E 70, 045101 (2004).
\bibitem{cressPRL} J.C. Cressoni, M.A.A. da Silva, G.M. Viswanathan, Phys. Rev. Lett. 98, 070603 (2007)
\bibitem{einstein} A. Einstein, Ann. Physik 17, 549 (1905). 
\bibitem{bach} L. Bachelier, Annales Scientifiques de l'Ecole Normale Superieure 17, 21 (1900). 
\bibitem{montroll} E. W. Montroll and G. H. Weiss, J. Math. Phys. 6, 167 (1965).
\bibitem{katja} D. Bedeaux, K. Lakatos-Lindenberg and K. Shuler, J. Math. Phys. 12, 2116 (1971); K. Lindenberg and R. Cukier, J. Chem. Phys. 62, 3271 (1975).
\bibitem{mike} \emph{L\'evy Flights and Related Topics in Physics}, ed. M.F. Shlesinger, G.M. Zaslavskii and U. Frisch (Springer, Berlin 1995).
\bibitem{vmkgme} V. M. Kenkre in \emph{Statistical Mechanics and Statistical Methods in Theory and Application} ed. U. Landman (New York, Plenum, September 1977), pp. 441-461.
\bibitem{grigo} P. Grigolini, Adv. Chem. Phys. 113, 357 (2006).
\bibitem{bruce} B.J. West, \emph{Mathematical Models as a Tool for the Social Sciences} (Taylor and Francis, London 1980).
\bibitem{metzler} R. Metzler and J. Klafter, Phys. Rep. 339, 1 (2000).
\bibitem{mantegna} R.N. Mantegna and H.E. Stanley, \emph{An Introduction to Econophysics} (Cambridge University Press, Cambridge, 2000).
\bibitem{gandhi} G.M. Viswanathan, V. Afanasyev, S.V. Buldyrev, E.J. Murphy, P.A. Prince, and H.E. Stanley, Nature (London) 381, 413 (1996).
\bibitem{theorem} See, e.g., I. Gy\"ori and G. Ladas, \emph{Oscillation Theory of Delay Differential Equations with Applications}, (Clarendon Press, Oxford, 1991). 
\bibitem{notte} Because $x_0$ and $t_0$ were not available from Ref. \cite{cressPRL} they were adjusted but only once per curve.
\bibitem{joneskenkre} D. Jones and V. M. Kenkre, in preparation.
\bibitem{sevillakenkre} F. Sevilla and V. M. Kenkre, J. Phys. Condens. Matter 19, 065113 (2007); see also V. M. Kenkre, E. Fukushima and D. Sheltraw, J. Magn. Reson. 128, 62 (1997).
\bibitem{note1} In keeping with earlier work, we take  both these quantities to be infinitesimal but their ratio to be finite. The normalizing time is then $t$ as in the original formulation.
\bibitem{note2} Our $D\eta(t)$ corresponds to $tD(t)$ in the notation of Ref. \cite{st}.
\end{references}
\end{document}